\documentclass[journal]{IEEEtran}
\usepackage{ucs}
\usepackage[utf8x]{inputenc}
\usepackage[cmex10]{amsmath}
\usepackage{cite, amsfonts, amssymb, amsthm, bm, bbm, graphicx, relsize, multirow, booktabs, tikz,subfigure,soul}
\usepackage[american]{babel}
\usepackage[T1]{fontenc}
\usepackage{algorithmic, algorithm}
\usepackage[multiple]{footmisc}
\theoremstyle{definition}

\theoremstyle{remark}

\renewcommand{\IEEEQED}{\IEEEQEDopen}

\begin{document}

\title{Multiuser Millimeter Wave MIMO Channel Estimation with Hybrid Beamforming}
\author{Stefano Buzzi, {\em Senior Member}, {\em IEEE}, and Carmen D'Andrea
\thanks{The authors are with the Department of Electrical and Information Engineering, University of Cassino and Lazio Meridionale, I-03043 Cassino, Italy (buzzi@unicas.it, carmen.dandrea@unicas.it).}
}
\maketitle
\pagestyle{empty}
\begin{abstract}
This paper focuses on  multiuser MIMO channel estimation and data transmission at  millimeter wave (mmWave) 
frequencies. The proposed approach relies on the time-division-duplex (TDD) protocol and is based on two distinct phases. First of all, the Base Station (BS) sends a suitable probing signal so that all the Mobile Stations (MSs), using a subspace tracking algorithm, can estimate the dominant left singular vectors of their BS-to-MS propagation channel. Then, each MS, using the estimated dominant left singular vectors as pre-coding beamformers, sends a suitable pilot sequence so that the BS can estimate the corresponding  right dominant channel singular vectors and the corresponding eigenvalues. 
The low-complexity projection approximation subspace tracking with deflation (PASTd) algorithm is used at the MSs for dominant subspace estimation, while pilot-matched (PM) and zero-forcing (ZF) reception is used at the BS. The proposed algorithms can be used in conjuction with an analog RF beamformer and are shown to exhibit very good performance. 
\end{abstract}


\section{Introduction and system model}

The use of frequency bands in the range 10−100 GHz, a.k.a. millimeter waves (mmWaves), for cellular communications, is among the most striking technological innovations brought by fifth generation (5G) wireless networks \cite{whatwillbe}. 
Indeed, the scarcity of available frequency bands in the sub-6 GHz spectrum has been the main thrust for considering the use of higher frequencies for cellular applications, and indeed recent research \cite{itwillwork} has shown that mmWaves, despite increased path-loss and atmospheric absorption phenomena, can be actually used for cellular communications over short-range distances (up to 100-200 meters), provided that multiple antennas are used at both sides of the communication link: MIMO processing, thus, is one distinguishing and key feature of mmWave systems.

This paper is devoted to the problem of joint multiuser MIMO channel estimation and data transmission in a single-cell  wireless network using mmWave frequencies and the TDD protocol. Building upon the recent work \cite{WSA2017_channel_estimation}, wherein the issue of subspace-based single-user MIMO channel estimation at mmWave frequencies is tackled, we develop a framework wherein first the BS sends a suitable probing signal in order to let the $K$ MSs estimate - using a subspace tracking algorithm - the dominant left channel eigenvectors; then, the MSs, using the estimated vectors as pre-coding beamformers, send pilot sequences to enable channel estimation at the BS. 

We denote by $N_{\rm BS}$ the number of antennas at the BS, and by $N_{\rm MS}$ the number of antennas at the MSs, and assume for simplicity a bi-dimensional model, i.e. both the BS and the MSs are equipped with a uniform linear array (ULA). We denote by $M$ the multiplexing order.

We denote by $\mathbf{H}_k$ the $(N_{\rm MS} \times N_{\rm BS})$-dimensional matrix representing the  channel
from the BS to the $k$-th MS. Due to  TDD operation the reverse-link propagation channel is expressed as $\mathbf{H}_k^H$. The popular clustered mmWave channel model is used \cite{buzzidandreachannel_model}, i.e. the channel results from single-bounce reflections from cluster of scatterers with random locations. 
Beamforming at the BS and at the MSs is of the hybrid type; we will denote by 
$N_{\rm BS}^{\rm RF}<N_{\rm BS}$ and $N_{\rm MS}^{\rm RF}<N_{\rm MS}$  the number of RF chains at the BS and at each MS, respectively.
The front-end processing both at the BS and at the MSs consists of an analog RF combining matrix aimed at reducing the number of RF chains needed to implement the base-band processing. From a mathematical point of view, 
the beamforming matrices at the $k$-th MS and at the BS can be expressed as
\begin{equation}
\begin{array}{llll}
\mathbf{D}_{\rm k}&= & \mathbf{D}_{k,{\rm RF}}\mathbf{D}_{k, {\rm BB}} \; , \qquad \mbox{and} \\
\mathbf{D}_{\rm BS}&= & \mathbf{D}_{\rm BS,RF}\mathbf{D}_{\rm BS,BB} \; ,
\end{array}
\label{eq:HYcombiners}
\end{equation}
respectively. In \eqref{eq:HYcombiners}, 
 $\mathbf{D}_{k,{\rm RF}}$ is an $(N_{\rm MS} \times N_{\rm MS}^{\rm RF})$-dimensional matrix with unit-norm entries, while  $\mathbf{D}_{k,{\rm BB}}$ is an $(N_{\rm MS}^{\rm RF} \times M)$-dimensional matrix with no constraint on its entries. Similarly,  $\mathbf{D}_{\rm BS,RF}$ is an $(N_{\rm BS} \times N_{\rm BS}^{\rm RF})$-dimensional matrix with unit-norm entries, and $\mathbf{D}_{\rm BS,BB}$ is an $(N_{\rm BS}^{\rm RF} \times M)$-dimensional baseband combining matrix. 
For the sake of simplicity, we assume that $\mathbf{D}_{k,{\rm RF}}$ and $\mathbf{D}_{\rm BS,RF}$ have a fixed structure and in particular contain on their column the ULA array responses corresponding to a grid of discrete angles uniformly spanning the range $[-\pi/2, \pi/2]$. 
More precisely, letting $\mathbf{a}_N(\theta)$ denote the $N$-element unit-norm ULA array response corresponding to the angle $\theta$, the matrix $\mathbf{D}_{k,{\rm RF}}$ contains on its columns the vectors 
$\mathbf{a}_{N_{\rm MS}}(\theta_i^{\rm MS})$, with  
$\theta_i^{\rm MS}=-\frac{\pi}{2} + \pi \frac{(i-1)}{N_{\rm MS}^{\rm RF}}$, with $i=1, \ldots, N_{\rm MS}^{\rm RF}$.
A similar definition can be given for the BS analog beamformer $\mathbf{D}_{\rm BS,RF}$, containing on its columns the vectors 
$\mathbf{a}_{N_{\rm BS}}(\theta_i^{\rm BS})$, with  
$\theta_i^{\rm BS}=-\frac{\pi}{2} + \pi \frac{(i-1)}{N_{\rm BS}^{\rm RF}}$, with $i=1, \ldots, N_{\rm BS}^{\rm RF}$.
Now, the  cascade of the BS analog beamformer, the channel $\mathbf{H}_k$ and the $k$-th MS analog beamformer can be modeled through the matrix  $\widetilde{\mathbf{H}}_k=\mathbf{D}^H_{k,{\rm RF}} \mathbf{H}_k \mathbf{D}_{\rm BS,RF}$, of dimension 
$N_{\rm MS}^{\rm RF} \times N_{\rm BS}^{\rm RF}$.  As a consequence, the channel estimation schemes outlined in the sequel will be applied to the composite channels $\widetilde{\mathbf{H}}_k$, with $k=1, \ldots, K$. Notice also that 
letting $N_{\rm BS}^{\rm RF}=N_{\rm BS}$ and $N_{\rm MS}^{\rm RF}=N_{\rm MS}$  and taking the analog beamformers equal to an identity matrix the procedures developed in the following describe a system with fully-digital  beamforming (FD-BF).

\section{Channel estimation and data communication}

The proposed protocol for channel estimation consists of two successive phases. 

\subsection{Phase (a): Subspace-based dominant eigenvectors estimation at the MSs}
In  this phase, the BS transmits a suitable probing signal and the MSs estimate the dominant left eigenvectors of their respective BS-to-MS channel matrix. Let  $\mathbf{s}_{\rm BS}(n)$, with $n=1, \ldots, P_{\rm BS}$, be a sequence of $N_{\rm BS}^{\rm RF}$-dimensional 
random column 
vectors with identity covariance matrix\footnote{As an example, a sequence of random uniform binary-valued antipodal symbols can be used.}. These vectors  are transmitted by the BS at (discrete) time $n=1, \ldots, P_{\rm BS}$ with power $P_T$; the signal received at the $k$-th MS at time $n$ is expressed as
the following $(N_{\rm MS}^{\rm RF} \times 1)$-dimensional vector:

\begin{equation}
\mathbf{r}_{k}(n)=\sqrt{\frac{P_T}{{\rm tr}(\mathbf{D}_{\rm BS, RF}\mathbf{D}_{\rm BS, RF}^H)}}\widetilde{\mathbf{H}}_k\mathbf{s}_{\rm BS}(n)+\mathbf{w}_{k}(n) \; ,
\label{eq:receivedMS}
\end{equation} 
where $\mathbf{w}_{k}(n)$ is the $N_{\rm MS}^{\rm RF}$-dimensional AWGN vector, modeled as a sequence of $\mathcal{CN} (0, \sigma^2_n \mathbf{D}_{k, {\rm RF}}^H\mathbf{D}_{k, {\rm RF}})$ independent random vectors.
Letting  $\widetilde{\mathbf{H}}_k=\mathbf{U}_k\mathbf{\Lambda}_k\mathbf{V}_k^H$ denote the singular value decomposition of the $k$-th channel matrix,  the covariance matrix of the received signal at the $k$-th MS is expressed as
\begin{equation}
\mathbf{R}_{k}=E\left[\mathbf{r}_{k}(n)\mathbf{r}_{k}^H(n)\right]=\mathbf{U}_k\mathbf{\Lambda}_k^2
\mathbf{U}_k^H+ \sigma^2_n \mathbf{D}_{k, {\rm RF}}^H\mathbf{D}_{k, {\rm RF}} \; .
\label{eq:covMS}
\end{equation}
Given \eqref{eq:covMS}, and assuming that $\mathbf{D}_{k, {\rm RF}}^H\mathbf{D}_{k, {\rm RF}}$ can be approximated with the identity matrix\footnote{Actually we can dismiss this approximation through the use of a whitening filter, although we are not following this steps here for the sake of simplicity.}  it is thus easily seen that the $k$-th MS can estimate the $M$ dominant left singular vectors of the channel matrix by estimating the $M$ dominant directions of the subspace spanned by the received vectors 
$\mathbf{r}_{k}(n)$, with $n=1, \ldots, P_{\rm BS}$. The PASTd Algorithm \ref{PASTd},  reported in \cite{yang1995projection}, can be used at each MS for this task.
Let now
$\mathbf{D}_{k,{\rm BB}}$ be the $(N_{\rm MS}^{\rm RF} \times M)$-dimensional matrix containing the estimate of the $M$ dominant singular vectors of $\mathbf{R}_{k}$. It is worth noting that the illustrated processing does not require that the MSs have knowledge of the random data sequence $\mathbf{s}_{\rm BS}(n)$ sent by the BS. 

\subsection{Phase (b): Uplink channel estimation at the BS}
After $P_{\rm BS}$ symbol intervals, phase (a) is over and phase (b) starts. Let us now denote by $\mathbf{\Phi}_k$ an $(M \times P_{\rm MS})$-dimensional matrix containing on its rows the $M$  unit-energy pilot sequences assigned to user $k$; we assume that the rows of  $\mathbf{\Phi}_k$ are orthogonal, while no orthogonality is required for pilot sequences assigned to different users. In particular, we will use in the numerical simulations random binary pilot sequences with the constraint that the rows of each matrix $\mathbf{\Phi}_k$ be orthogonal.
The generic $k$-th MS transmits, over $P_{\rm MS}$ consecutive signaling slots, the  matrix 
$\sqrt{\alpha_k} \mathbf{D}_{k,{\rm BB}} \Phi_k$, with
$\alpha_k$  a proper coefficient ruling the power transmitted by the $k$-th MS. The signal received at the BS can be thus expressed as the following 
$(N_{\rm BS}^{\rm RF} \times P_{\rm MS})$-dimensional matrix:
\begin{equation}
\begin{array}{lll}
\mathbf{Y}&\! = \! \displaystyle \sum_{k=1}^K \sqrt{\alpha_k}\widetilde{\mathbf{H}}_k^H \mathbf{D}_{k,{\rm BB}} \Phi_k \!+ \mathbf{Z} 
& \!\approx \!\displaystyle \sum_{k=1}^K \sum_{i=1}^M\! \lambda_{k,i}\mathbf{v}_{k,i}\Phi_k(i,:) \!+ \mathbf{Z}
\end{array}
\label{eq:received_BS}
\end{equation}
wherein  $\mathbf{Z}$ contains the thermal noise contribution, we have assumed, with no loss of generality, that the diagonal entries of $\mathbf{\Lambda}_k$ are ordered according to a decreasing magnitude, and $\lambda_{k,i}$ and $\mathbf{v}_{k,i}$ represent the $i$-th diagonal entry of $\sqrt{\alpha_k} \mathbf{\Lambda}_k$ and the $i$-th column of $\mathbf{V}_k$, respectively.
Now, based on observable \eqref{eq:received_BS}, and relying on the knowledge of the pilot matrices $\Phi_k$, a number of algorithms can be envisaged to estimate the dominant right channel eigenvectors $\mathbf{v}_{k,i}$. 
A simple algorithm relies on PM filtering, i.e. the product $\lambda_{k,i} \mathbf{v}_{k,i}$ can be estimated as follows:
$\widehat{\lambda_{k,i}  \mathbf{v}}_{k,i}=\mathbf{Y}[\mathbf{\Phi}_k(i,:)]^H$.
Alternatively, if  $P_{\rm MS}\geq MK$,  a ZF approach can be also used; in particular, the estimate of the matrix
$[\lambda_{k,1} \mathbf{v}_{k,1}, \ldots,  \lambda_{k,M} \mathbf{v}_{k,M}]$ if formed by considering the statistic $\mathbf{Y}\mathbf{Z}_k$, where the $(P_{\rm MS} \times M)$-dimensional matrix $\mathbf{Z}_k$ is such that $\mathbf{\Phi}_k \mathbf{Z}_k=\mathbf{I}_M$ and $\mathbf{\Phi}_j  \mathbf{Z}_k$ is the all-zero matrix for all $j \neq k$. 

\subsection{The data communication phase}
Once phases (a) and (b) are over, each MS has an estimate of the $M$ dominant left eigenvectors of the BS-to-MS channel and the BS has an estimate of the matrix $[\lambda_{k,1} \mathbf{v}_{k,1}, \ldots,  \lambda_{k,M} \mathbf{v}_{k,M}]$, for all $k=1, \ldots, K$. Note also that coherent channel information is available only at the BS, since the MSs have estimated the dominand channel eigendirections with no knowledge on the signal sent from the BS. 
Based on the available knowledge, data communication can take place. Regarding the uplink, the generic $k$-th MS can use as transmit beamformer the estimated matrix $\mathbf{D}_{k,{\rm BB}}$, while the BS can simply use as receiver beamforming for the $k$-th MS the available estimate of the matrix  $[\lambda_{k,1} \mathbf{v}_{k,1}, \ldots,  \lambda_{k,M} \mathbf{v}_{k,M}]$. Alternative reception schemes can be conceived, e.g., ZF or linear minimum mean square error receivers, but they are omitted here due to lack of space. 

Regarding the downlink, again the BS can use the available estimate of the matrix  $[\lambda_{k,1} \mathbf{v}_{k,1}, \ldots,  \lambda_{k,M} \mathbf{v}_{k,M}]$ as the $k$-th user transmit beamformer, while $\mathbf{D}_{k,{\rm BB}}$ is used as receiver beamformer at the $k$-th MS. Also in this case more sophisticated precoding schemes can be used at the BS, although they are not treated here.

\begin{algorithm}[!t]

\caption{The PASTd Algorithm}

\begin{algorithmic}[1]

\label{PASTd}

\STATE $\mathbf{x}_1(n)=\mathbf{r}(n)$

\FOR { $m=1:M$}

\STATE  {$y_m(n)=\mathbf{u}_m^H(n-1)\mathbf{x}_m(n)$}

\STATE {$\lambda_m(n)=\beta\lambda_m(n-1)+|y_m(n)|^2$}

\STATE {$\mathbf{u}_m(n)\! \! = \! \! \mathbf{u}_m(\!n-1\!)\!+\!\left[\mathbf{x}_m(n)\!-\!\mathbf{u}_m(n-1)y_m(n)\right]\frac{y_m(n)^*}{\lambda_m(n)}$}

\STATE $\mathbf{x}_{m+1}(n)=\mathbf{x}_m(n)-\mathbf{u}_m(n)y_m(n)$

\ENDFOR

\end{algorithmic}

\end{algorithm}

\begin{figure}[t]
\centering
\includegraphics[scale=0.25]{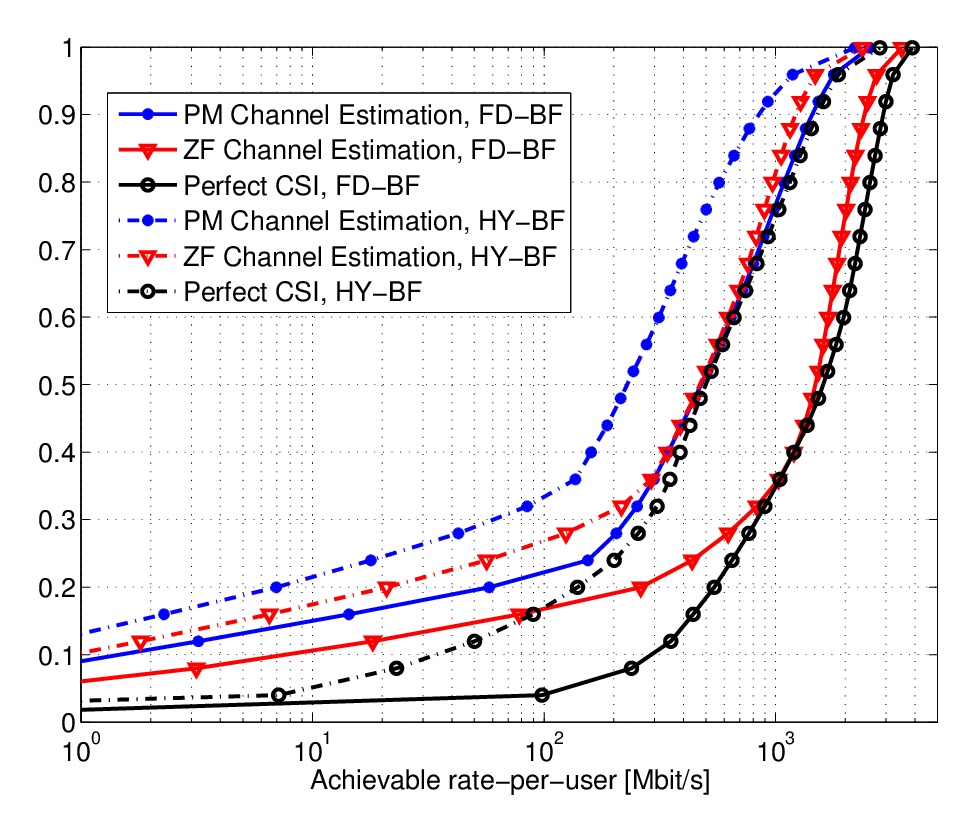}
\caption{CDF of the downlink achievable rate-per-user for a system with $K=5$ users. We considered a system with $N_{\rm BS}=64$ and $N_{MS}=4$; for the HY-BF case we have used $N_{\rm BS}^{\rm RF}=16$ and  $N_{\rm MS}^{\rm RF}=2$. The multiplexing order is $M=1$.}
\label{Fig:DL}
\end{figure}

\begin{figure}[t]
\centering
\includegraphics[scale=0.25]{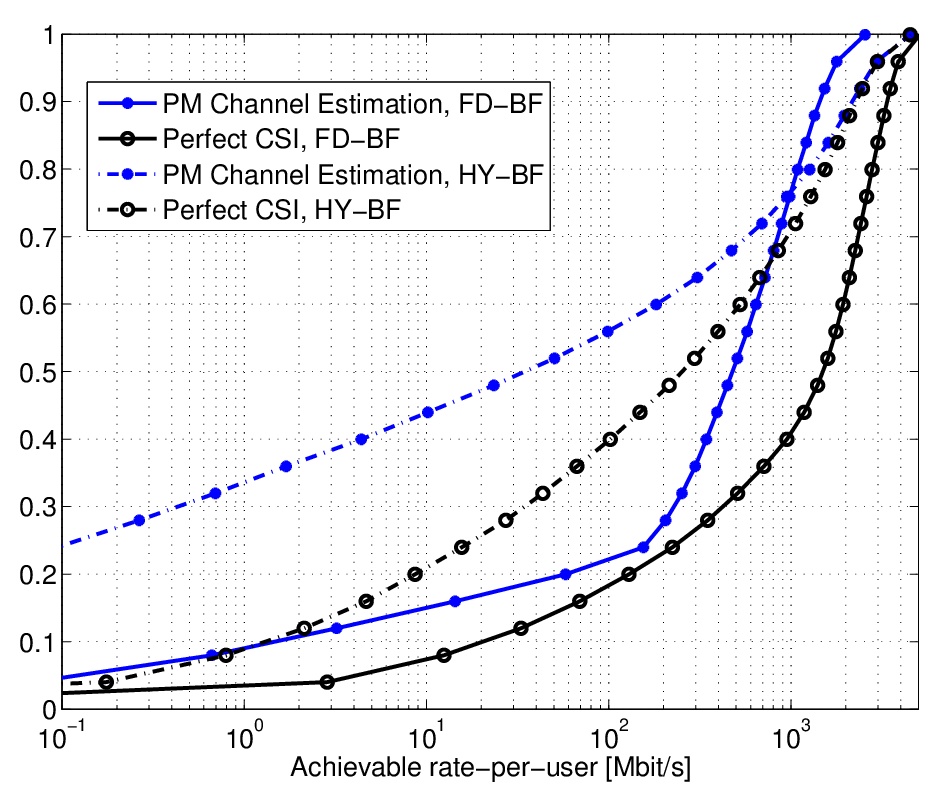}
\caption{CDF of the uplink achievable rate-per-user. The considered parameters are the same as those of Fig. 1.}
\label{Fig:UL}
\end{figure}
\section{Numerical Results}
We consider as performance measure the achievable rate expressed through the usual log-det formula.
We considered a system with one BS, $K=5$ MSs, and a communication bandwidth of $W = 500$ MHz centered
over the carrier frequency $f_0=73$ GHz. The distance between the BS and the MSs was randomly chosen in the range $[5, 100]$ m; the additive thermal noise is assumed to have a power spectral density of $-174$ dBm/Hz, while the front-end receiver at the BS and at the MSs is assumed to have a noise figure of $6$ dB. During phase (a) the BS transmit power is 1 W and the pilot length is $P_{\rm BS}=60$; in phase (b), the MSs transmit with a power of 0.1 W, and the pilot length is $P_{\rm MS}=32$. During the data communication phase the BS and MSs transmit powers are again 1 W and 0. 1W, respectively.
The shown results come from an average over 5000 random scenario realizations with independent channels. 
Fig. 1 shows the CDF of the downlink achievable  rate-per-user of the system, while Fig. 2 shows the CDF of the uplink 
achievable rate-per-user. We consider both the case of FD-BF and of hybrid beamforming 
(HY-BF). The considered system parameters are included in the figures' captions. With regard to Fig. 1 we consider both the cases in which the BS beamformer is obtained using PM channel estimation and ZF channel estimation; the case of perfect channel state information (CSI) is also reported. With regard to Fig. 2, instead, we consider the cases of channel-matched beamforming at the BS using either the PM channel estimated or the perfect CSI\footnote{As already discussed, more sophisticated precoding strategies can be readily applied.}. Results are quite satisfactory and encouraging. Focusing  on the median rate, we see from Fig. 1 that the HY-BF structure with PM channel estimation achieves a rate-per-user of 243 Mbit/s; using ZF processing at the receiver the median rate-per-user increases to 500 Mbit/s and, interestingly, it is just $5.4 \%$ smaller than the median rate for the case of perfect CSI. Remarkably, the top-10$\%$ luckiest MSs can achieve an individual rate of 1.28 Gbit/s for the case of ZF channel estimation. 
Inspecting Fig. 2, similar conclusions can be drawn for the uplink, although the rates are generally lower than for the downlink (recall that the MSs transmit with 10dB less power than the BS). As an instance, the uplink median achievable rate-per-user is, for the HY-BF case, 50 Mbit/s with BS decoding based on PM channel estimates; if FD-HY is used, this number has a 10-fold increase up to 507 Mbit/s. 
Other numerical results, not shown here due to lack of space, have shown that the proposed approach works well also with a larger number of users. In particular, for the case of $K=15$ users, the downlink median achievable rate-per-user with ZF channel estimation at the BS and HY-BF is 192 Mbit/s, i.e. the reduction factor for the rate-per-user is $500/192=2.6$ despite the fact that the number of MSs has been increased by a factor of 3.

\section{Conclusion}
This paper has been focused on the problem of multiuser channel estimation and data communication for a single-cell wireless MIMO links at mmWave frequencies. The proposed approach takes into account hardware complexity constraints by using a simple and fized analog RF beamformer, and has been shown to achieve satisfactory performance.

\bibliographystyle{IEEEtran}

\bibliography{FracProg_SB,finalRefs,references}

\end{document}